\documentclass[a4paper, amsfonts, amssymb, amsmath, reprint, showkeys, nofootinbib, twoside]{revtex4-1}
\usepackage[english]{babel}
\usepackage[utf8]{inputenc}
\usepackage[colorinlistoftodos, color=green!40, prependcaption]{todonotes}
\usepackage{amsthm}
\usepackage{xcolor}
\usepackage{enumitem}
\usepackage{graphicx}
\usepackage[T1]{fontenc}
\usepackage[pdftex, pdftitle={Article}, pdfauthor={Author}]{hyperref} 
\begin{document}
\title{Emulating dark energy models with known equation of state via the created cold dark matter scenario}

\author{V\'ictor H. C\'ardenas$^1$}
\email{victor.cardenas@uv.cl}
\author{Miguel Cruz$^2$}
\email{miguelcruz02@uv.mx}
\affiliation{$^1$Instituto de F\'{\i}sica y Astronom\'ia, Universidad de Valpara\'iso, Gran Breta\~na 1111, Valpara\'iso, Chile\\
$^2$Facultad de F\'{\i}sica, Universidad Veracruzana 91097, Xalapa, Veracruz, M\'exico}

\date{\today} 

\begin{abstract}
In this work we establish only at background dynamics level, the equivalence between the Created Cold Dark Matter (CCDM) model and diverse dark energy (DE) models. We find that for a barotropic or linear equation of state (EoS) for the DE, $p=w \rho$, and standard matter sector the corresponding CCDM model is described by the same functional structure of the ratio of particle production, $\Gamma$. For a different EoS the functional form of the $\Gamma$ term is not longer maintained, however, in the case of a polytropic EoS given by the Chaplygin gas the resulting $\Gamma$ term can be written as the one obtained in the barotropic case under certain considerations.
\end{abstract}

\keywords{matter creation, dark energy, FLRW}

\maketitle

\section{Introduction}
One of the most mysterious components in the universe budget is DE. This component is not only invisible except for their large scale effect on the universe expansion but also is the dominant part of the total mass energy of the observable universe. This represents a challenge for our conventional understanding of gravity and cosmological dynamics at late times \cite{Frieman:2008sn, Li:2012dt, bookDE}. Besides, the enigmatic nature of this elusive component is further compounded by the intriguing obtained results when the fundamental processes of particle physics are used to describe it, see for instance Ref. \cite{phande}. In the framework of General Relativity the simplest DE model introduces a cosmological constant, namely $\Lambda$, leading to the so far successful $\Lambda$CDM model, which have passed many observational tests. Although appealing based on a better fit to observations, the nature of the cosmological constant and its particular value are still as mysterious as their own existence. Additionally, in this era of improvements in the collection and precision of astrophysical data, we have more questions about the DE dominated universe instead of a clear exit light sign for this paradigm \cite{riess2024}; everything seems to indicate that we require a look beyond the $\Lambda$CDM model, interesting recent results (and related to some DE models studied here) regarding this point can be found in \cite{vagnozzi23, vagnozzi24}. However, the walk on this swampy road could become easier if we complement the description of the gravitational interaction at large scales with fundamental physical principles. This latter possibility is explored in this work.\\ 

The history of the cosmic evolution can be strongly influenced by the quantum mechanism of particle production in the early universe \cite{creation, davies}. The impact of these production processes on the cosmological dynamics began to be studied from a phenomenological point of view some decades ago. An interesting result of these studies is the equivalence found between the particle production scheme and the bulk viscous scenario \cite{classic0}, later extended to the so-called adiabatic matter creation framework \cite{classic, classic2}. The findings within this analogy between both models can be summarized as follows: bulk viscous cosmologies are associated with a generalization of ideal fluids, in this context the model requires of constitutive equations to describe viscous pressures that can be interpreted as small non-equilibrium contributions for the energy-momentum tensor \cite{maartens}, in general these contributions lead to a negative effective pressure which is expected to be responsible of the accelerated cosmic expansion. For matter creation also a negative effective pressure arises and can be associated with this effect since the dynamics of the model takes into account the non-conservation of the number of particles in the fluid, in this scheme such condition plays the role of the constitutive equations in bulk viscous cosmologies. Therefore both scenarios have a richer structure from the thermodynamics point of view than those cosmological models where perfect and non-interacting fluids are used. Besides, for the case of matter creation there are not restrictions with respect to the near equilibrium condition as with a negative bulk viscous pressure, thus such description can be implemented more consistently than bulk viscous cosmologies for an expanding fluid, where the near equilibrium condition could be violated \cite{nonlinear}.

The main result of this work at background dynamics level is to show that under certain conditions (which we will specify below) a wide variety of well-known cosmological models can be obtained from a single fluid description in which matter creation is allowed. It is important to mention that in the matter creation scenario in addition to the geometrical state of the universe and the usual physical variables that describe the cosmic fluid, pressure and energy density; we also have to consider the particle density, $n$, this simple extension in the description of the fluid leads to a reinterpretation of Einstein equations and irreversible processes. In this sense, the background equivalence between models established in this work does not prevail at perturbations level. However, an interesting feature of this scenario is that the particle production term, $\Gamma$, can be reconstructed from the cosmological parameters of the mimicked model, this construction eliminates the arbitrariness regarding this $\Gamma$-term since it is usually imposed by hand in this kind of framework. Besides such term maintains a specific structure that depends explicitly on the parameter state of the DE sector. We expect that from the connection provided between matter creation and different cosmic scenarios, cosmological models that are known to resolve some of the currently tensions in cosmology could be mapped straightforwardly to a single fluid scheme in which is possible to have a new and more robust description of the content of the universe from the thermodynamics point of view. This could provide a novel panorama to exploit in the cosmological context at background level for a plethora of DE models known so far. An interesting work in which the mimicry between different DE models is explored in great detail can be found in Ref. \cite{bamba}.\\ 

The paper is organized as follows. In Section \ref{sec:single} we provide a brief description and summarize the main results of the CCDM model in the context of a Friedmann-Lemaitre-Robertson-Walker (FLRW) spacetime. Section \ref{sec:analog} is devoted to the analogy at background level between the CCDM model and some well-known DE models with constant parameter state, $w$, by reconstructing the particle production rate. The case with variable parameter state is also discussed. In Section \ref{sec:poly} we discuss the Chaplygin EoS and some of its generalizations. We show that under an approximated barotropic EoS obtained from the Chaplygin scenario the form of the particle production rate coincides with one obtained in the usual DE models. In Section \ref{sec:final} we give the final comments of our work. 

\section{Matter creation: a single component}
\label{sec:single}
Assuming valid the cosmological principle we use a flat FLRW metric
\begin{equation}
    ds^2 = -c^2dt^{2} + a(t)^2(dx^2+dy^2+dz^2),
\end{equation}
where $a(t)$ is the scale factor. Because matter is being created, the vector particle number $N^{\mu}=nu^{\mu}$ is not conserved so $N^{\mu}_{;\mu}=n\Gamma$, considering the FLRW metric the previous equation takes the form \cite{classic0, classic, classic2}
\begin{equation}\label{ene}
  \dot{n} + 3 H n = n\Gamma,
\end{equation}
where $n$ is the number density then $n:=N/V$ with $N$ representing the total number of particles contained in the volume $V$, $u^{\mu}$ is the 4-velocity, and $\Gamma$ is the particle creation rate. 

The vector entropy flux being defined by $S^{\mu}=\sigma N^{\mu}=\sigma n u^{\mu}$ satisfy the second law for an isolated system
\begin{equation}
    S^{\mu}_{; \mu}=n\dot{\sigma} + n\sigma \Gamma \geq 0. \label{eq:flux}
\end{equation}
where $\sigma$ is the specific entropy. We can also write this in terms of the entropy density $s=\sigma n$, then $S^{\mu}=su^{\mu}$ and the second law reads as $\dot{s} + 3Hs \geq 0$. Clearly if the specific entropy is constant ($\dot{\sigma}=0$) and there is no particle creation ($\Gamma =0$) then we recover $S^{\mu}_{; \mu}=0$ for equilibrium. However, under the aforementioned adiabatic condition positive entropy production is still guaranteed for $\Gamma>0$.

The usual thermodynamic expression $dU=dQ-pdV$, which establishes the conservation of the internal energy, $U=\rho V$, can be written in terms of $n$ as
\begin{equation}
    d\left(\frac{\rho}{n}\right) = dq -pd\left(\frac{1}{n}\right),
\end{equation}
where $dQ$ is the heat received by the system during some time interval and we have defined $dq:=(dQ)/N$. If we consider $N$ as a time dependent function, the generalization to open systems of the energy conservation equation given previously takes the form,
\begin{equation}\label{firstlaw0}
d(\rho V)+ pdV -(h/n)d(nV)=dQ,
\end{equation}
with $h=(\rho+p)$ defined as the enthalpy (per unit volume). For adiabatic transformations the expression given above is simply  
\begin{equation}\label{firstlaw}
d(\rho V)+ pdV -(h/n)d(nV)=0.
\end{equation}
In such a transformation, the change in the number of particles can be interpreted as the {\it heat} received by the system. In the cosmological scenario this change is entirely due to the transfer of energy from gravitation to matter. For a constant number of particles and adiabatic transformations, we recover the usual expression $d(\rho V)+pdV =0$. The important thing to stress here is that the universe evolution continues to be adiabatic, in the sense that the entropy per particle remains unchanged ($\dot{s}=0$) but the entropy of the fluid changes due to the increasing number of produced particles, this can be seen from the second law of thermodynamics, decomposing the entropy change as
\begin{equation}
    dS = d_{e}S+d_{c}S \geq 0, \label{eq:total0}
\end{equation}
where $d_{e}S$ is an entropy flow and $d_{c}S$ is the creation entropy. Due to the change in the number of particles, $N$, the total differential of the entropy can be written also in terms of the chemical potential denoted as $\mu$ \cite{callen} 
\begin{equation}
    Td(sV) = d(\rho V)+pdV-\mu d(nV), \label{eq:total}
\end{equation}
with $s=S/V$ and $\mu \geq 0$. For closed systems and adiabatic transformations we obtain directly the condition $dS=0 \rightarrow d_{c}S=0$. If we now turn on the matter creation contribution we get from Eqs. (\ref{firstlaw}), (\ref{eq:total0}) and (\ref{eq:total})
\begin{equation}
    TdS=Td_{c}S = (h/n)d(nV) - \mu d(nV) \geq 0,
\end{equation}
which in turn results as
\begin{equation}\label{second}
dS=\frac{s}{n}d(nV) > 0,
\end{equation}
by means of the Euler relation $h=Ts+\mu n$. Notice that the time derivative of the latter expression depends explicitly on $\Gamma$ due to Eq. (\ref{ene}). According to the second law (\ref{second}), variations in the number of particles satisfying, $dN=d(nV) > 0$ are allowed. This means that in the context of open systems and irreversible processes, spacetime can produce matter whereas the reverse process is forbidden. The appearance of particles in this context is due to the existence of fluctuations in the geometrical background associated to quantum fluctuations of matter fields, in other words, the energy of produced particles is obtained from the gravitational field; see Ref. \cite{creation} for a complete analysis. Under these conditions an extra contribution can be interpreted in (\ref{firstlaw}) as a non thermal pressure defined as
\begin{equation}\label{pc}
p_c=-\left(\frac{\rho + p}{3H}\right) \Gamma.
\end{equation}
This is the source that produces the acceleration of the universe expansion. Once the particle number increases with the volume, we obtain a negative pressure \cite{classic0, classic, classic2}.

From (\ref{firstlaw}) we find that the new set of Einstein equations are
\begin{equation}\label{friedman}
  H^2 + \frac{k}{a^2} =  \frac{8\pi G}{3} \rho
\end{equation}
\begin{equation}\label{rho}
  \dot{\rho} = \frac{\dot{n}}{n}(\rho + p)
\end{equation}
and the previously derived relation (\ref{ene}). The set of
equations (\ref{ene}), (\ref{friedman}) and (\ref{rho}) completely
specified the system evolution. It is also useful to combine
(\ref{pc}) with (\ref{friedman}) and (\ref{rho}) to eliminate $\rho$
and obtain
\begin{equation}\label{eq5}
2\frac{\ddot{a}}{a}+\frac{\dot{a}^2}{a^2}+\frac{k}{a^2}=-8 \pi G(p +
p_c).
\end{equation}
The standard adiabatic evolution is easily recovered: setting
$\Gamma = 0$ implies that $\dot{n}/n=-3H$, which leads to the usual conservation equation from (\ref{rho}). A class of de Sitter
solution is obtained with $\dot{n}=\dot{\rho}=0$ and arbitrary
pressure $p$. Moreover, there exist another class of solutions where Eq. (\ref{rho}) enable us to determine the pressure; for example if $\rho=m n$, with $m$ being a constant characterizing the mass of the created particle, Eq. (\ref{rho}) implies $p=0$, and furthermore if $\rho=aT^4$ and $n=bT^3$ implies $p=\rho/3$.
\section{The matter creation analog model: background dynamics}
\label{sec:analog}
In the scheme of the matter creation scenario, we can construct for any DE model and standard matter sector, an analog matter creation one. This can be done considering only the Eqs. (\ref{ene}) and (\ref{rho}) assuming $p=0$ for the created matter, we obtain
\begin{equation}
    \dot{\rho} + 3H\rho = \rho \Gamma,
\end{equation}
which can be written as
\begin{equation}\label{eq: gammadef}
    \frac{d}{dt}\log \left( a^3 \rho \right) = \Gamma.
\end{equation}
In the next sub sections we study in detail different DE models and their matter creation analogs.

\subsection{$\Lambda$CDM model}

The simplest and so far the more successful model that describes the observational data is the Lambda ($\Lambda$) cold dark matter model.

In this case, the total matter density $\rho$ must consider to be {\it both} dark matter and the cosmological constant contribution, so $\rho = \rho_m^0 a^{-3} + \rho_{\Lambda}$ with $\rho_{\Lambda}$ being a constant and the equation (\ref{eq: gammadef}) leads to
\begin{equation}\label{eq: glcdm}
    \Gamma = \frac{d}{dt}\log \left( \rho_m^0 + a^3 \rho_{\Lambda} \right) = \frac{3H\rho_{\Lambda}}{\rho_m^0a^{-3}+ \rho_{\Lambda}}=\frac{3H\rho_{\Lambda}}{\rho}.
\end{equation}
In fact, using this particle creation rate we obtain for the creation pressure (\ref{pc}) with $p=0$,
\begin{equation}
    p_c = -\frac{\rho}{3H}\Gamma = - \rho_{\Lambda}, 
    \label{eq:pressurel}
\end{equation}
which identifies the creation pressure with the cosmological constant equation of state. 

So, the standard model of DM plus a cosmological constant is equivalent -- at background level -- to a matter creation model with a single pressureless component and a particle creation rate given by (\ref{eq: glcdm}). 

\subsection{The wCDM model}

Another model that is tested and used extensively in the literature is the $w$CDM that consist in a DE component with an explicit EoS parameter $w$ which is assumed to be constant \cite{wcdm}. In this case, the total energy density is given by
\begin{equation}
    \rho = \rho_m^0 a^{-3} + \rho_x^0 a^{-3(1+w)}.
\end{equation}
Following our procedure as before, we get a rate of particle production as
\begin{equation}\label{eq: gwcdm}
    \Gamma =  -\frac{3Hw \rho_x^{0}a^{3(1+w)}}{\rho} = -\frac{3Hw\rho_x}{\rho},
\end{equation}
produces an evolution identical to the one for the $w$CDM model. This results coincides with (\ref{eq: glcdm}) in the limit of $w=-1$ as it should be. Notice that the positivity of the source term (\ref{eq: gwcdm}) is guaranteed.

\subsection{The dark interaction model}

In this subsection we would like to study the dark interaction model, for a review of this cosmological model and its scopes see \cite{interact} . This is also equivalent to a model for a dynamical EoS $w(a)$, which are indistinguishable from each other. Let us assume a model for DE where $\rho = \rho_m + \rho_x$ and each component satisfy the relations
\begin{eqnarray}
\dot{\rho}_m +3H\rho_m &=& Q,\label{eq: romq} \\
\dot{\rho}_x +3H(1+w)\rho_x &=& -Q,\label{eq: roxq}
\end{eqnarray}
with an {\it arbitrary} $Q$. From (\ref{eq: gammadef}) we get
\begin{equation}
    \Gamma = \frac{d}{dt} \log \left( a^3 \rho_m + a^3\rho_x \right).
\end{equation}
Taking the time derivative of the latter expression and using Eqs. (\ref{eq: romq}) and (\ref{eq: roxq}) we get
\begin{equation}\label{eq: ggral}
\Gamma = -\frac{ 3 H w \rho_x}{\rho},
\end{equation}
which is again of the same form of (\ref{eq: glcdm}) and (\ref{eq: gwcdm}). This means that with (\ref{eq: ggral}) the matter creation scenario enables to account for interacting DE models with an arbitrary interaction term $Q$.  

\subsection{A variable EoS parameter}

Everything we have studied in previous sections, specially the main result which is the form that takes the particle creation rate $\Gamma$ in Eq. (\ref{eq: ggral}) has been based on assuming a constant EoS parameter for the DE component. In this section we explore the general case for a variable EoS parameter $w(a)$.

The idea is to answer if this case corresponds to just replacing $w(a)$ into the Eq. (\ref{eq: gammadef}). We will assume an arbitrary function of the scale factor for the parameter state of the DE fluid, $w = w(a)$, in this case we can write 
\begin{equation}
    \dot{\rho}_{x} + 3H[1+w(a)]\rho_{x} = 0,
\end{equation}
then one gets
\begin{equation}
    \rho_{x} = \rho^{0}_{x}e^{-3\int^{a}_{1}\frac{1+w(a')}{a'}da' },
    \label{eq:generalde}
\end{equation}
where we have considered $a_{0} = 1$, i.e., unity is the present time value for the scale factor. The total energy density in this case is given by
\begin{equation}
    \rho = \rho^{0}_{m}a^{-3} + \rho^{0}_{x}e^{-3\int^{a}_{1}\frac{1+w(a')}{a'}da' },
\end{equation}
which is inserted into the equation (\ref{eq: gammadef}), yielding again
\begin{equation}
    \Gamma = -\frac{3H\rho_{x}}{\rho}w(a).
\end{equation}
So this form of $\Gamma$ is robust for any DE component with a barotropic EoS.

\section{A polytropic EoS}
\label{sec:poly}

As discussed above, the structure of the source term in the analog matter creation model for different DE models seems to be robust for a linear relation between the energy density and pressure of the DE fluid, $p_{x} \propto \rho_{x}$. In this section we discuss a more general case for the EoS given by a polytropic form. Let us assume the cosmological case of the Chaplygin gas as dominant component of the universe with EoS given by \cite{chapli1}
\begin{equation}
    p=-\frac{A}{\rho},
    \label{eq:eoschaply}
\end{equation}
with $A$ a positive constant. In this case the continuity equation for the energy density is given as
\begin{equation}
    \dot{\rho} + 3H(\rho+p)=0,
    \label{eq:contipoly}
\end{equation}
assuming that $\rho$ as a function of time can be represented as a function of the scale factor $\rho(a)$, the conservation law can be integrated in a straightforward manner using Eq. (\ref{eq:eoschaply}), yielding
\begin{equation}
    \rho(a) = \sqrt{A+\frac{B}{a^{6}}},
    \label{eq:chaply1}
\end{equation}
where $B$ is an integration constant. Notice that for positive $B$ and small scale factor one gets
\begin{equation}
    \rho(a) \approx \frac{\sqrt{B}}{a^{3}},
\end{equation}
and for large values of $a$
\begin{equation}
    \rho(a) \approx \sqrt{A}\ \ \longrightarrow \ \ p \approx -\sqrt{A}.
\end{equation}
Then the matter given in Eq. (\ref{eq:chaply1}) behaves like dust at the beginning of cosmological evolution and at the end like a cosmological constant. If we implement the procedure described before and we insert Eq. (\ref{eq:chaply1}) into (\ref{eq: gammadef}) we have 
\begin{equation}
    \Gamma = -\frac{3Hp}{\rho}.
\end{equation}
As can be seen, in this case the form of the source term is more simple than the cased discussed before but its structure differs from the one associated to linear models. However, if we consider the subleading terms emerging from (\ref{eq:chaply1}) for large scale factor one gets
\begin{equation}
    \rho(a) \approx \sqrt{A} + \frac{B}{\sqrt{4A}}a^{-6},
    \label{eq:sub1}
\end{equation}
which according to Eq. (\ref{eq:eoschaply}) leads to
\begin{equation}
    p(a) \approx -\sqrt{A} + \frac{B}{\sqrt{4A}}a^{-6}.
\label{eq:sub2}
\end{equation}
Notice that Eqs. (\ref{eq:sub1}) and (\ref{eq:sub2}) represent a cosmological constant model and matter sector, the EoS for each component are $p_{m}=\rho_{m}$ and $p_{x} = -\rho_{x}$, i.e., both contributions to the energy density are described with a barotropic EoS respectively. For this scenario the source term takes the form
\begin{equation}
    \Gamma = \frac{3H\rho_{x}}{\rho}-\frac{3H\rho_{m}}{\rho},
    \label{eq:gammagen}
\end{equation}
which resembles the structure discussed previously for different dark energy models. This form of $\Gamma$ was obtained by inserting Eq. (\ref{eq:sub1}) into (\ref{eq: gammadef}). The first term of the latter expression corresponds to the cosmological constant contribution and the second term appears since now we are dealing with a stiff matter sector that evolves as $a^{-6}$ instead $a^{-3}$ and parameter state given by $w=1$. Regarding this point it is important to mention that in this case the structure of the source term differs from the one discussed in the previous section for different DE models since for all those models the total pressure of the fluid is characterized simply by the DE contribution, then from the usual conservation equation for the total energy density, $\dot{\rho}+3H(\rho+p_{x})=0$, we can associate to the fluid the following parameter state
\begin{equation}
    w=w_{x}\frac{\rho_{x}}{\rho},
\end{equation}
under the assumption of linear relationship between $\rho_{x}$ and $p_{x}$. The total pressure of the cosmic fluid is characterized by the parameter state of only one of the components of the universe, this was confirmed for the $\Lambda$CDM model and discussed below Eq. (\ref{eq:pressurel}). Notice that according to Eq. (\ref{eq:sub2}), the total pressure of the Chaplygin gas has two contributions.

On the other hand, since stiff matter is associated to early cosmic evolution, we can also consider small deviations from the usual behavior for the matter sector by extending our previous results by means of the generalized Chaplygin EoS in order to accommodate late times evolution, in this case it is more convenient the following expression \cite{chapli}
\begin{equation}
  p=-\frac{A}{\rho^{\alpha}},
    \label{eq:eoschaply1}  
\end{equation}
with $\alpha$ being a constant parameter given in the interval $0\leq \alpha \leq 1$. The energy density takes the form
\begin{equation}
\rho(a) = \left[A+\frac{B}{a^{3(1+\alpha)}} \right]^{\frac{1}{1+\alpha}}.
\end{equation}
where $B$ is again an integration constant. If we expand for large scale factor the density given above we have \cite{chapli}
\begin{equation}
    \rho(a) \approx A^{1/(1+\alpha)} + \frac{1}{1+\alpha}\frac{BA^{-\alpha/(1+\alpha)}}{a^{3(1+\alpha)}},
    \label{eq:sub3}
\end{equation}
then from the EoS (\ref{eq:eoschaply1}) we write
\begin{equation}
    p(a) \approx -A^{1/(1+\alpha)} + \frac{\alpha}{1+\alpha}\frac{BA^{-\alpha/(1+\alpha)}}{a^{3(1+\alpha)}},
    \label{eq:sub4}
\end{equation}
in this case we have a mixture of a cosmological constant contribution and a modified matter sector described by the EoS $p_{m} = \alpha \rho_{m}$, which leads to the source term
\begin{equation}
    \Gamma = \frac{3H\rho_{x}}{\rho}-\frac{3H\rho_{m}}{\rho}\alpha.
\label{eq:finsour}
\end{equation}
It is worthy to mention that for $\alpha = 0$ the matter sector described by Eq. (\ref{eq:sub3}) scales in the usual form as function of the scale factor, $a^{-3}$, thus the source term (\ref{eq:finsour}) coincides with Eq. (\ref{eq: glcdm}) since the total pressure (\ref{eq:sub4}) is characterized by a single term associated to the cosmological constant. Notice that at late times the positivity of (\ref{eq:finsour}) is guaranteed due to the domination of DE, therefore
\begin{equation}
    \Gamma \approx 3H\left[1-(1+\alpha)\frac{\rho_{m}}{\rho_{x}} \right] \approx 3H,
\label{eq:finsour1}
\end{equation}
this value for $\Gamma$ is commonly proposed as an Ansatz in the literature \cite{3h}. Finally we would like to comment that the structure obtained for the source term as given in Eqs. (\ref{eq:gammagen}) and (\ref{eq:finsour}) is also maintained by the EoS \cite{chapli2}
\begin{equation}
    p=A\rho-\frac{B}{\rho^{\alpha}},
    \label{eq:eoslast}
\end{equation}
in this case the DE sector is described by a constant energy density with EoS $p_{x}=[A-(1+A)^{\alpha}]\rho_{x}$ and for the modified matter sector we have, $p_{m} = [\alpha + A(1+\alpha)]\rho_{m}$ with scaling behavior given as $a^{-3(1+\alpha)(1+A)}$, i.e., the barotropic structure for each specie prevails. If we consider $A=0$ in Eq. (\ref{eq:eoslast}) we recover the results discussed for the EoS given in Eq. (\ref{eq:eoschaply1}): $p_{x}=-\rho_{x}$ and $p_{m}=\alpha \rho_{m}$. The structure of the results discussed above also holds for an alternative generalization of the Chaplygin's model as the one introduced in \cite{chaplirengo}.

\section{Final remarks}
\label{sec:final}
In this work we established a mapping only at background dynamics level between different DE models and the CCDM approach through the rate of particles production  
\begin{equation}
    \Gamma = -\frac{3H\rho_{x}}{\rho}w(a),
    \label{eq:finalG}
\end{equation}
this expression will be valid as long as there is a linear (or barotropic) relationship between density and pressure, $p_{x}(a) = w(a)\rho_{x}(a)$; where $\rho_{x}$ is the energy density of the DE model of interest, $w(a)$ is the parameter state and $\rho$ is the total energy density of the cosmological model, $\rho(a) = \rho_{m}(a)+\rho_{x}(a)$, being $\rho_{m}$ the energy density of the matter sector with dust behavior, i.e., pressureless fluid which in general is used to describe the dark matter content in the universe. An interesting feature of this model is the scope that the single component emerging from the production scenario described by (\ref{eq:finalG}) now has. At first place is able to reproduce the components of the dark sector and in consequence is viable to characterize the background dynamics for the late times cosmic evolution of these models. Besides, the structure (\ref{eq:finalG}) remains valid for any DE model with constant and/or variable parameter state, $w$. This approach represents a richer scenario from the thermodynamics point of view for any DE model since the entropy production is turned on by means of the source term $\Gamma$ as can be seen in (\ref{second}), the cosmic expansion is now an irrerversible process. On the other hand, this new cosmological approach can be applied to a wide variety of DE models, see for instance Ref. \cite{jaber}, where with a single parametrization for $w(a)$ and a linear relation for $\rho$ and $p$ the authors can reproduce the generic behavior of the most widely used DE models for accelerated expansion with infrared corrections. See also \cite{odintsovw} where an {\it inhomogeneous} EoS, i.e., $w_{x}=w_{x}(\rho_{x},H,\dot{H})$, can lead to cosmological models which are identical to some usual DE models with modified parameter state but where the linear structure holds for the EoS.\\ 

For the case of a polytropic EoS given by the Chaplygin gas and some of its generalizations, we observe that the generic form (\ref{eq:finalG}) is not maintained. However, as commented previously, under certain limit the Chaplygin model describes a single component whose energy content behaves as cosmological constant plus a modified matter sector with linear EoS for each component, we observe that in such case the structure (\ref{eq:finalG}) is now splitted out in two terms due to the contribution of the modified matter sector to the total pressure of the cosmic fluid, in this case we have
\begin{equation}
    \Gamma = \frac{3H\rho_{x}}{\rho}-\frac{3H\rho_{m}}{\rho}w_{m},
\end{equation}
which resembles the form given in (\ref{eq:finalG}) with $w=-1$ and $w_{m}=0$. In all cases discussed for the generalized Chaplygin gas we also observed that if the matter sector scales in the usual form we have $w_{m} = 0$, therefore the structure given in (\ref{eq: glcdm}) for the LCDM model is recovered. Thus the linear relationship between pressure and energy density for the components of the dark sector is partially responsible of the generic form found for the source term in the CCDM model. We must also take into account the form of the total pressure of the fluid, extra contributions to pressure determines the number of terms appearing in $\Gamma$. This second approach can be of interest in models as unimodular gravity, where the matter sector is modified and the DE content is described by the cosmological constant, see for instance \cite{unimodular}.\\ 

It is worthy to mention that despite the relevance that some DE models have in solving issues as the cosmological coincidence problem or certain inconsistencies with some astrophysical data sets by admitting a negative energy density for the DE sector at some stage of the cosmic evolution are not allowed in this approach since they lead to $\Gamma<0$, thus violating the second law of thermodynamics in this scenario at first glance, see Ref. \cite{second} for this kind of DE models. We leave this subject open for deeper exploration.

Finally, we emphasize that there are still a lot of subtle issues worth investigating when the perturbative analysis of this new kind of matter creation scenario is implemented. Due to the form of the constructed $\Gamma$-term some differences are expected to be found from the usual models considered in the literature for $\Gamma$. However, we feel that any further exploration along this line would justify a separate analysis. We will report this elsewhere.

\section*{Acknowledgments}
M. Cruz acknowledges the warm hospitality of the Physics and Astronomy Institute of Valparaiso University where this work was done. This work was partially supported by S.N.I.I. (CONAHCyT-M\'exico) (MC).

\end{document}